\documentclass[aps, prl, twocolumn, superscriptaddress, showpacs]{revtex4}

\usepackage{graphicx}

\begin{document}

\title{Experimental Realization of Maximum Confidence State Discrimination for the Extraction of Quantum Information}
\author{Peter J. Mosley}
\affiliation{Clarendon Laboratory, University of Oxford, Parks Rd, Oxford, OX1 3PU, United Kingdom}
\author{Sarah Croke}
\email{sarah@phys.strath.ac.uk}
\affiliation{Department of Physics, University of Strathclyde, Glasgow, G4 0NG, United Kingdom}
\affiliation{Department of Mathematics, University of Glasgow, Glasgow, G12 8QW, United Kingdom}
\author{Ian A. Walmsley}
\affiliation{Clarendon Laboratory, University of Oxford, Parks Rd, Oxford, OX1 3PU, United Kingdom}
\author{Stephen M. Barnett}
\affiliation{Department of Physics, University of Strathclyde, Glasgow, G4 0NG, United Kingdom}

\pacs{03.65.Ta, 03.67.Hk, 42.50.Xa}

\newcommand{\bra}[1]{\langle #1|}
\newcommand{\ket}[1]{|#1\rangle}
\newcommand{\braket}[2]{\langle #1|#2\rangle}

\begin{abstract}
We present the first experimental demonstration of the maximum confidence measurement strategy for quantum state discrimination. Applying this strategy to an arbitrary set of states assigns to each input state a measurement outcome which, when realized, gives the highest possible confidence that the state was indeed present. The theoretically optimal measurement for discriminating between three equiprobable symmetric qubit states is implemented in a polarization-based free-space interferometer. The maximum confidence in the measurement result is 2/3. This is the first explicit demonstration that an improvement in the confidence over the optimal minimum error measurement is possible for linearly dependent states.
\end{abstract}

\maketitle

There is no fundamental difficulty in extracting information encoded in orthogonal quantum states, for example the horizontal and vertical polarization modes of a photon. In principle, by projecting onto orthogonal states spanning the Hilbert space, the state can be determined with certainty, with the number of possible measurement outcomes equal to the number of input states. However, for a set of \textit{n} nonorthogonal states, it is no longer possible to write down \textit{n} positive operators each of which have a non-zero overlap with only one state and which together form a complete measurement. Therefore, the input state can never be determined perfectly.

This intrinsic inability of any measurement to distinguish between nonorthogonal quantum states with certainty has led to the development of a suite of strategies optimizing various figures of merit relating to this inevitable uncertainty \cite{chefles}. The most straightforward criterion for constructing a measurement of this type is to minimize the probability of identifying the state incorrectly as a result of measurement \cite{hel,hol,yuen}. The optimal minimum error measurement, however, is generally difficult to find, and is known only for certain special cases. Some of these have been implemented experimentally using the polarization of light as a two-level system \cite{riis,cla2}. Alternatively, for linearly independent states, one can design a measurement which can determine the initial state without error, but an additional, inconclusive outcome is necessary which, when obtained, gives limited or no information about the state of the system \cite{unamb, chefles2, jae, peres}.

Unambiguous discrimination between two pure states has been demonstrated in both fiber and free-space optical assemblies \cite{hutt, cla1}, using optical polarization. Unambiguous discrimination has also been demonstrated between both pure and mixed states in three dimensions using a multi-rail interferometer to create a multi-level optical system \cite{moh}. A further strategy, which maximizes the mutual information shared by the transmitting and receiving parties \cite{davies, sas} has also been realized experimentally \cite{cla2, miz}.

In this paper we present an experimental implementation of an optimal maximum confidence measurement. This measurement maximizes our confidence in identifying any given state in a set, i.e. when the measurement outcome leads us to identify a particular state, the probability that this state was really present is maximized \cite{maxconf}. The experiment was performed for three equiprobable, symmetric, linearly dependent input states, using optical polarization as a two-level system, and linear optical elements to create and manipulate the states.

It is possible to describe any physically realizable measurement using a probability operator measure (POM \cite{hel}, or positive operator valued measure -- POVM \cite{peres2}) consisting of elements $\{ \hat{\Pi}_i \}$ with corresponding measurement outcomes $\{ \omega_i \}$. These elements are required to be non-negative and to form a complete set. For a system in state $\hat{\rho}$, each outcome $\omega_j$ is then obtained with probability $\mathrm{Tr}(\hat{\rho}\hat{\Pi}_j)$. In the state discrimination problem, a measurement is made to distinguish between a known set of states $\{\hat{\rho}_i \}$, occurring with \textit{a priori} probabilities $\{ p_i \}$, and it is usually assumed that outcome $\omega_j$ leads us to hypothesize that the state of the system was $\hat{\rho}_j$. The probability that state $\hat{\rho}_j$ was indeed present can then be written using Bayes' Rule as:
\begin{equation}
\label{prob_correct}
P(\hat{\rho}_j | \omega_j) = \frac{P(\hat{\rho}_j) P(\omega_j | \hat{\rho}_j)}{P(\omega_j)} = \frac{p_j \mathrm{Tr}(\hat{\rho}_j \hat{\Pi}_j)}{\mathrm{Tr}(\hat{\rho} \hat{\Pi}_j)},
\end{equation}
where $\hat \rho = \sum_i p_i \hat \rho_i$. The maximum confidence measurement maximizes this quantity for each given input state $\hat{\rho}_j$ and for pure states, $\ket{\Psi_j}$, has POM elements given by \cite{maxconf, kos}:
\begin{equation}
\label{pies}
\hat{\Pi}_j \propto \hat{\rho}^{-1} \hat{\Psi}_j  \hat{\rho}^{-1}.
\end{equation}
It can be seen that within this strategy, as with unambiguous discrimination, we still have some freedom in how we construct the measurement, manifested in the choice of the constants of proportionality.  An inconclusive outcome is also sometimes needed to form a complete measurement.

The aim of our experiment was to demonstrate maximum confidence discrimination between the pure states described by the kets:
\begin{eqnarray}
\label{states}
\ket{\Psi_0} & = &\cos\theta\ket{0} + \sin\theta\ket{1}, \nonumber \\
\ket{\Psi_1} & = & \cos\theta\ket{0} + e^{2 \pi i / 3} \sin\theta\ket{1}, \\
\ket{\Psi_2} & = & \cos\theta\ket{0} + e^{-2 \pi i / 3} \sin\theta\ket{1}, \nonumber
\end{eqnarray}
where $0^\circ \leq \theta \leq 45^\circ$ and $\ket{0}, \ket{1}$ are orthonormal basis kets (in our implementation these correspond to right and left circular polarization -- $\ket{R}, \ket{L}$ -- respectively).  We chose to implement the maximum confidence measurement for this set, described in \cite{maxconf}, where the constants of proportionality are chosen to minimize the probability of occurrence of the inconclusive result. This measurement is described by the POM with elements $\{\hat{\Pi}_i = (3 \cos^{2} \theta)^{-1} \ket{\phi_i}\bra{\phi_i}, \hat{\Pi}_? = (1 - \tan^{2} \theta) \ket{0} \bra{0} \}$ where
\begin{eqnarray}
\label{pom_elements}
\ket{\phi_0} & =  &\sin\theta\ket{0} + \cos\theta\ket{1}, \nonumber \\
\ket{\phi_1} & = & \sin\theta\ket{0} + e^{2 \pi i / 3} \cos\theta\ket{1}, \\
\ket{\phi_2} & = & \sin\theta\ket{0} + e^{-2 \pi i / 3} \cos\theta\ket{1}. \nonumber
\end{eqnarray}
and $\hat{\Pi_{?}}$ corresponds to the inconclusive result.  For these POM elements the confidence that the input state really was $\ket{\Psi_i}$ when outcome $\omega_i$ is obtained is 2/3.

\begin{figure}
\includegraphics[width=70truemm]{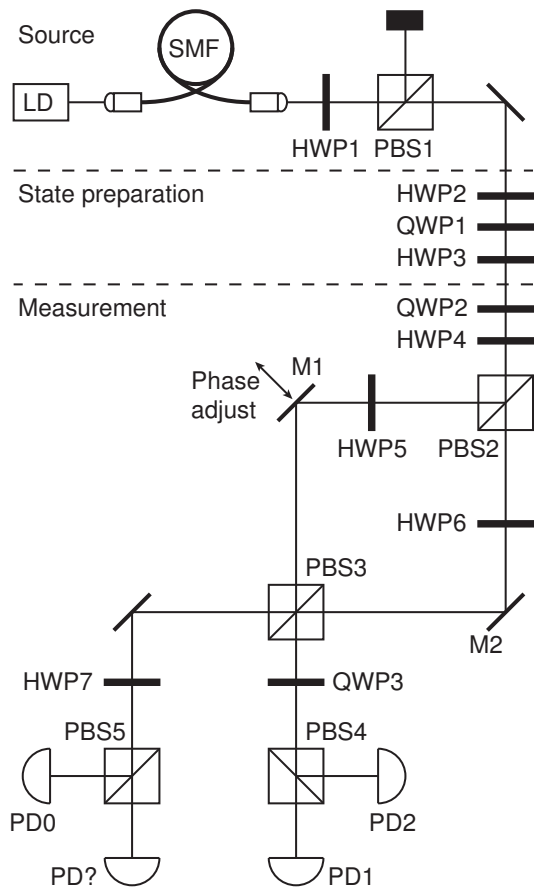}
\caption{Experimental apparatus. LD=Laser diode, SMF=single mode fiber, PBS1--5=polarizing beamsplitters, HWP1--7=zero-order half waveplates, QWP1--3=zero-order quarter waveplates, PD0--2, PD?=amplified photodiodes}
\label{app}
\end{figure}

Our experimental apparatus consisted of three principal sections -- the light source, followed by preparation of the input state, and finally the measurement -- as shown in Fig.\ \ref{app}. The source was a CW 810nm laser diode with a FWHM bandwidth of 0.88nm. This was coupled into a single mode fiber to act as a spatial filter, the output of which was collimated using an aspheric lens. The beam was initially sent through a half waveplate (HWP1) followed by a polarizing beamsplitter (PBS1) to provide clean horizontal input polarization and to act as a variable attenuator.  All the PBSs transmitted horizontally polarized light ($\ket{H}$) and reflected vertical polarization ($\ket{V}$) and all the waveplates were zero-order.

For state preparation, HWP2 was oriented at $\theta/2 -45^{\circ}$, and HWP3 at $-\phi/4$ where $\phi = 0, 120^{\circ}, -120^{\circ}$ for $\ket{\Psi_0}, \ket{\Psi_1}$ and $\ket{\Psi_2}$ respectively.  QWP1 was oriented at $-45^{\circ}$ for all input states.  Waveplate angle conventions were as in \cite{cla2}.

\begin{figure*}
\includegraphics[width=170truemm]{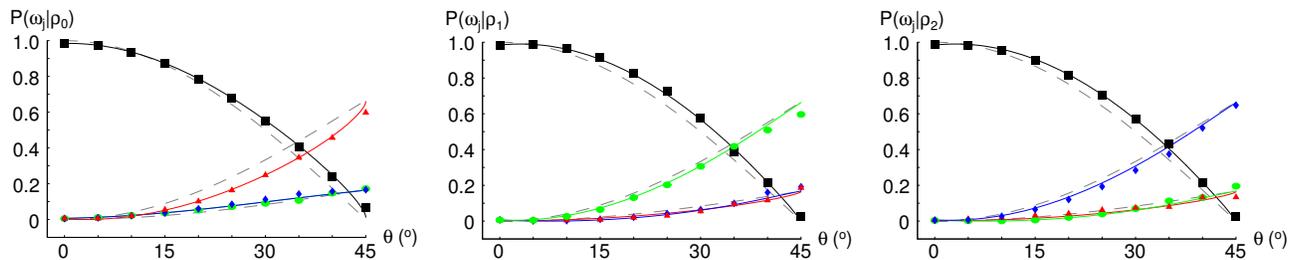}
\caption{(Color online) Experimental results alongside the theoretical predictions without error (dashed lines) and with non-ideal beamsplitters PBS2 and PBS3 (colored lines), as a function of $\theta$, for each of the input states $\ket{\Psi_0}, \ket{\Psi_1}, \ket{\Psi_2}$ (l-r).  Points show the normalized voltage at each detector, with red triangles, black squares, green circles, blue diamonds corresponding to PD0, PD?, PD1, PD2 respectively.  Experimental errors in the measured voltages are smaller than the size of the data points.}
\label{data}
\end{figure*}

The state discrimination section performs the maximum confidence measurement outlined above, described by the POM $\{ \hat{\Pi}_0, \hat{\Pi}_1 , \hat{\Pi}_2, \hat{\Pi}_? \}$, on each of the input states in turn.  The vacuum input to the interferometer provides an additional two orthogonal field modes, thus giving the four orthogonal modes needed in order to realize a four-outcome measurement \cite{naimark, peres2}.  Our experimental design groups together the four outputs in pairs so that two orthogonal modes in one output arm of the interferometer correspond to the inconclusive result and the result $\omega_0$, while two in the other arm correspond to results $\omega_1$ and $\omega_2$.  The appropriate unitary transformations, effected by HWP7 and QWP3, ensure that these modes can then be separated by means of PBS4 and PBS5.  All four outcomes of the measurement can therefore be accessed, and are monitored by photodetectors PD0-2 and PD?.  Full details of the experimental design are straightforward but lengthy, and will be discussed elsewhere.

The interferometer was aligned in a Mach-Zehnder configuration by setting the input polarization to 45$^\circ$ and both HWP5 and HWP6 to 0$^\circ$. The angle of M2 and position of M1 were adjusted to optimize the interference fringes in the two output ports. Mirror M1 was mounted on a precision translation stage, allowing the relative phase between the arms of the interferometer to be accurately varied. Altering the position of this mirror obviously changed the overlap of the recombining beams at PBS3, however, it was ascertained that the phase could be adjusted over a range greater than 2$\pi$ radians without affecting the quality of the interference fringes. This was monitored using the output voltages from detectors PD0 and PD? with HWP7 set to rotate the polarization by 45$^\circ$ back to the $\ket{H} \ket{V}$ basis. The visibility of the interference was always greater than 99\% after alignment. The interferometer was encased in a box to achieve phase stability for at least 10 minutes. The phase could be controlled from outside the box, minimizing any disturbance to the interferometer.

The photodetectors used were amplified photodiodes (Thorlabs PDA520-EC) whose linearity and relative calibration were measured over the relevant range of incident power to be within 0.1\% and 2.5\% respectively. The output voltages of the four photodetectors were displayed and recorded using a digital oscilloscope (LeCroy Wavepro 7100). HWP1 was rotated to attenuate the power in the transmitted arm of PBS1 to keep the photodetectors within their linear response range.  All the waveplates used were calibrated to within 0.1$^\circ$ and were measured to have extinction ratios of better than 1:2000. The waveplates were held in precision mounts that allowed the angles of the optic axes to be set to within 0.1$^\circ$ with a repeatability of 0.1$^\circ$. The extinction ratios of the beamsplitters were measured to be around 1:200.

The maximum confidence measurements were performed for ten values of the parameter $\theta$, ranging from $0^\circ$ to $45^\circ$ in $5^\circ$ steps.  At each value of $\theta$, the optic axes of the waveplates in the measurement section were first set to the appropriate orientations.  For HWP4-7, these are given in Table \ref{WPsettings}. QWP2 and QWP3 were both oriented at $45^{\circ}$ for all $\theta$.  Due to the symmetry of the measurement, for each given input state, two of the outputs theoretically have the same probability of occurrence; this was utilized to set the phase of the interferometer. With the input in state $\ket{\Psi_1}$ the phase was set by adjusting the position of M1 to minimize the difference between the outputs at detectors PD0 and PD2. Once set, it was possible to cycle quickly through the three input states by rotating HWP3. The need to move only one waveplate, coupled with the long-term stability of the interferometer, ensured phase stability between the three measurements for each value of $\theta$. The output voltages were recorded for 10 seconds and averaged to give each data point. For each input state, the voltages were normalized by dividing by the total voltage at all the detectors, giving the probability that input state $\ket{\Psi_i}$ results in outcome $\omega_j$.

\begin{table}
\begin{ruledtabular}
\begin{tabular}{ccccc}
$\theta$ & HWP4 & HWP5 & HWP6 & HWP7 \\
\hline
0 & 0 & 27.4 & 0 & 0 \\
5 & 1.3 & 27.4 & 1.8 & 2.2 \\
10 & 2.6 & 27.5 & 3.6 & 4.4 \\
15 & 4.0 & 27.8 & 5.4 & 6.9 \\
20 & 5.7 & 28.1 & 7.3 & 9.7 \\
25 & 7.7 & 28.7 & 9.3 & 12.9 \\
30 & 10.2 & 29.6 & 11.4 & 17.0 \\
35 & 13.5 & 31.2 & 13.6 & 22.2 \\
40 & 17.6 & 34.2 & 15.7 & 29.3 \\
45 & 22.5 & 45 & 17.6 & 45 \\
\end{tabular}
\end{ruledtabular}
\caption{Table showing the angle of the fast axes of HWP4-7 to the horizontal for each value of $\theta$.}
\label{WPsettings}
\end{table}

The normalized voltage measured at each detector is shown in Fig.\ \ref{data}.  Modeling of errors in different components showed that the largest influences are the non-ideal properties of PBS2 and PBS3. Fig.\ \ref{data} shows the predictions of the ideal theory and a non-ideal model in which 0.5\% of the incident light at PBS2 and PBS3 leaks into the ``wrong'' output port (consistent with the properties of the PBSs). The phase difference between the desired output and the erroneous component was left as a parameter in our model. The best agreement with the results was obtained when the transmission and reflection coefficients for these PBSs were all real. The small residual differences may be due to second order effects (waveplate errors and inaccuracies in the phase of the interferometer) and to the imperfect purity of the input states \cite{note}.

From these data we can calculate the confidence figure of merit -- the probability that when the measurement result leads us to identify state $\ket{\Psi_i}$, that state was indeed present. A voltage at detector PD$i, i=0,1,2$ indicates that the input state was $\ket{\Psi_i}$.  Thus if the normalized voltage at PD0, when the input state was $\ket{\Psi_i}$, is denoted V$_0^{(i)}$, the probability that a voltage at PD0 really was due to input state $\ket{\Psi_0}$ may be expressed in the form
\begin{equation}
\label{conf_out}
P(\rho_0 | \omega_0) = \frac{V_0^{(0)}(\theta)}{V_0^{(0)}(\theta) + V_0^{(1)}(\theta) + V_0^{(2)}(\theta)}.
\end{equation}
Similar expressions may be constructed for state $\ket{\Psi_1}$, $\ket{\Psi_2}$, all of which theoretically should be equal to 2/3.

\begin{figure}
\includegraphics[width=80truemm]{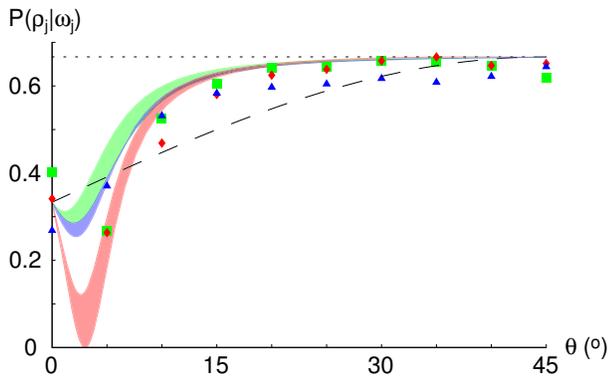}
\caption{(Color online) Experimental and theoretical values for the confidence figure of merit, and the confidence achieved by the optimal minimum error measurement (dashed line).  Red diamonds, green squares, blue triangles show the proportion of the voltage at detector PD$i, i=0,1,2$ respectively due to input state $\ket{\Psi_i}$, i.e. the confidence that a voltage in PD$i$ indicates that the input state was $\ket{\Psi_i}$.  Shaded regions show the range of values consistent with the non-ideal model, while the ideal theoretical value of 2/3 is also shown (dotted line).}
\label{confidence}
\end{figure}

Our results for this figure of merit are shown in Fig.\ \ref{confidence} alongside the ideal and non-ideal theory. The range of the non-ideal theory plots corresponds to the values of the PBS phase for which the non-ideal theory shows good agreement with the data in Fig.\ \ref{data}.  We can see from Fig.\ \ref{confidence} that the confidence measure is close to optimal for values of $\theta > 15^\circ$ but, as expected, this starts to drop as the angle between the the three input states becomes small. However, the advantage of the maximum confidence measurement strategy over that of minimum error is clear for states with $\theta < 35^\circ$. The dashed line in Fig.\ \ref{confidence} shows the confidence figure of merit for the optimal minimum error strategy; the values of the confidence calculated from our data are consistently greater for $10^\circ \leq \theta \leq 30^\circ$.

Although the measurement was performed using an attenuated coherent source, rather than single photons, because the outcomes are given simply by the outputs of single detectors they depend only on the second order correlation functions of the inputs. Therefore the results would not differ were the experiment to be repeated using a true single photon source and photon counting detectors, such as silicon avalanche photodiodes. The low noise characteristics of these units would allow an exact duplication of the results presented herein.

In conclusion, we have demonstrated an experimental implementation of a maximum confidence measurement strategy for discriminating between three nonorthogonal polarization states of light. The results of the experiment show a clear improvement in the confidence figure of merit over the optimal minimum error measurement for the same states.

This work was supported by the EPSRC (SC, SMB), the Synergy fund of the Universities of Glasgow and Strathclyde, the NSF (PJM, IAW), the DARPA QuIST program (IAW), and the EC under the Integrated QAP funded by the IST directorate as Contract No. 015848 (IAW). We thank E. Riis, J. Jeffers, E. Andersson, C. Gilson, R. Kosut, and H. Rabitz for helpful discussions.


\begin{thebibliography}{99}
\bibitem{note} The correlation time of the input laser beam was 2.5ps while the detector response time was 4$\mu$s.
\end{thebibliography}
\end{document}